\documentclass[draft]{agujournal2019}
\usepackage{url} 
\usepackage{lineno}
\usepackage[inline]{trackchanges} 
\usepackage{soul}

\draftfalse

\journalname{JGR: Space Physics}

\newcommand{\dalfven}{\Delta \lambda_{\mathrm{Alfv\acute{e}n}}}
\newcommand{\rj}{\textit{R}_{\mathrm{J}}}
\newcommand{\rgan}{\textit{R}_{\mathrm{Gan}}}
\newcommand{\ecar}{\textit{E}_{\mathrm{char}}}
\newcommand{\reur}{\textit{R}_{\mathrm{Eur}}}
\newcommand{\rio}{\textit{R}_{\mathrm{Io}}}

\begin{document}

%
%


\title{Properties of electrons accelerated by the Ganymede-magnetosphere interaction: survey of Juno high-latitude observations}

%
%




\authors{J. Rabia\affil{1}, V. Hue\affil{2}, N. André\affil{1,3}, Q. Nénon\affil{1}, J.R. Szalay\affil{4}, F. Allegrini\affil{5,6}, A.H. Sulaiman\affil{7}, C.K. Louis\affil{8}, T.K. Greathouse\affil{5}, Y. Sarkango\affil{4}, D. Santos-Costa\affil{5}, M. Blanc\affil{1,2}, E. Penou\affil{1}, P. Louarn\affil{1}, R.W. Ebert\affil{5,6}, G.R. Gladstone\affil{5,6}, A. Mura\affil{9}, J.E.P. Connerney\affil{10,11} and S.J. Bolton\affil{5}}


\affiliation{1}{Institut de Recherche en Astrophysique et Planétologie, CNRS-UPS-CNES, Toulouse, France}
\affiliation{2}{Aix-Marseille Université, CNRS-CNES, Institut Origines, LAM, Marseille, France}
\affiliation{3}{Institut Supérieur de l'Aéronautique et de l'Espace (ISAE-SUPAERO), Université de Toulouse, Toulouse, France}
\affiliation{4}{Department of Astrophysical Sciences, Princeton University, Princeton, NJ, USA}
\affiliation{5}{Southwest Research Institute, San Antonio, TX, USA}
\affiliation{6}{Department of Physics and Astronomy, University of Texas at San Antonio, San Antonio, TX, USA}
\affiliation{7}{School of Physics and Astronomy, Minnesota Institute for Astrophysics, University of Minnesota, Minneapolis, MN, USA,}
\affiliation{8}{LESIA, Observatoire de Paris, Université PSL, CNRS, Sorbonne Université, Université de Paris, Meudon, France}
\affiliation{9}{Institute for Space Astrophysics and Planetology, National Institute for Astrophysics, Rome, Italy}
\affiliation{10}{NASA Goddard Space Flight Center, Greenbelt, MD, USA}
\affiliation{11}{Space Research Corporation, Annapolis, MD, USA}





\correspondingauthor{J. Rabia}{jonas.rabia@irap.omp.eu}




\begin{keypoints}
\item Juno particle and UV measurements are combined with field-line tracing to identify 11 in situ crossings of the Ganymede flux tube. 
\item We provide a statistical study of the accelerated electrons observed in the high-latitude far-field region. 
\item We find two distinct regions in which the electrons properties, i.e., characteristic energy, energy flux, and distribution, greatly differ.
\end{keypoints}

%
%

%
%


\begin{abstract}

The encounter between the Jovian co-rotating plasma and Ganymede gives rise to electromagnetic waves that propagate along the magnetic field lines and accelerate particles by resonant or non-resonant wave-particle interaction. They ultimately precipitate into Jupiter's atmosphere and trigger auroral emissions. In this study, we use Juno/JADE, Juno/UVS data, and magnetic field line tracing to characterize the properties of electrons accelerated by the Ganymede-magnetosphere interaction in the far-field region. We show that the precipitating energy flux exhibits an exponential decay as a function of downtail distance from the moon, with an e-folding value of 29°, consistent with previous UV observations from the Hubble Space Telescope (HST). We characterize the electron energy distributions and show that two distributions exist. Electrons creating the Main Alfvén Wing (MAW) spot and the auroral tail always have broadband distribution and a mean characteristic energy of 2.2 keV while in the region connected to the Transhemispheric Electron Beam (TEB) spot the electrons are distributed non-monotonically, with a higher characteristic energy above 10 keV. Based on the observation of bidirectional electron beams, we suggest that Juno was located within the acceleration region during the 11 observations reported. We thus estimate that the acceleration region is extended, at least, between an altitude of 0.5 and 1.3 Jupiter radius above the 1-bar surface. Finally, we estimate the size of the interaction region in the Ganymede orbital plane using far-field measurements. These observations provide important insights for the study of particle acceleration processes involved in moon-magnetosphere interactions. 

\end{abstract}

\section*{Plain Language Summary}

The Galilean moons orbit in a plasma-rich environment, created by the intense volcanism of Io and transported radially outward in the Jovian magnetosphere. At the orbital locations of the moons, this plasma, co-rotating with Jupiter, flows at a velocity significantly higher than the moons' orbital speed. Consequently, the moons disturb the plasma flow. This interaction gives rise to a set of physical processes, including the generation of electromagnetic waves that propagate away from the moons and accelerate charged particles, triggering auroral emissions by precipitating into Jupiter's atmosphere. In this study, we investigate the properties of the electrons accelerated by the Ganymede-magnetosphere interaction. We use data from the JADE and UVS instruments onboard the Juno spacecraft as well as magnetic field line tracing methods. Following a statistical characterization of the electron properties, we compare our results with previous findings that have reported electron observations resulting from the Io- and Europa-magnetosphere interactions.

%
%

%


%
%
%
%

\section{Introduction}

The Juno mission \cite{bolton_2017}, in orbit around Jupiter since July 2016, is characterizing the auroral emissions and the accelerated particles that generate them with unprecedented temporal and spatial resolution. In particular, the study of the moons' auroral footprints and underlying physical processes is made possible by the highly-eccentric orbit of the spacecraft, which enables Juno to cross the different moons magnetic flux shells several times per orbit. In situ measurements obtained across the magnetic field lines connecting the moons orbital locations to Jupiter's atmosphere reveal the diversity of physical processes occurring during the moon-magnetosphere interactions.  

The most visible manifestation of these moon-magnetosphere interactions is the generation of UltraViolet (UV) and InfraRed (IR) auroral emissions at the feet of the connected magnetic field lines \cite{connerney_1993,prange_1996,clarke_1998,clarke_2002,mura_2017,mura_2018}. These auroral footprints are composed of a main spot, the Main Alfvén Wing (MAW) spot, that is linked to the respective moon Alfvén wings, followed by an auroral tail that extends in the direction of the magnetospheric plasma flow \cite{gerard_2006}. Depending on the location of the moons in the Jovian magnetodisc, a spot created by a Trans-hemispheric Electron Beam (TEB) can be observed upstream of the MAW spot or downstream of it \cite{bonfond_2008}. This TEB is created by electrons accelerated in the anti-jovian direction in one hemisphere, traveling along the magnetic field line and precipitating into the opposite hemisphere.    

The formation of these auroral structures is made possible by the generation in the moons local environments of electromagnetic waves that propagate away from the moons along the magnetic field lines as Alfvén waves, and accelerate particles. Such electromagnetic coupling between Jupiter and its moons was first predicted by \citeA{goldreich_1969}, and confirmed by Voyager 1 measurements at Io \cite{acuna_1981}. Theoretical formalisms were then developed to explain the propagation of Alfvén waves from the vicinity of the moons to Jupiter's auroral regions \cite{goertz_1980,neubaeur_1980,gurnett_1981,neubauer_1998}. More recently, these waves have been observed in situ by the Juno spacecraft \cite{sulaiman_2020}. Statistical studies have shown that the efficiency of the Alfvén waves-particles energy transfer is close to 10\% \cite{sulaiman_2023}, consistent with an interaction between electrons and Alfvén waves filamented by turbulent cascade \cite{hess_2010}. Theoretical and modeling studies have also been carried out to understand the propagation of Alfvén waves and their role in the particle acceleration in Jupiter's magnetosphere \cite{jacobsen_2007,saur_2018,damiano_2019,coffin_2022,lysak_2023}, showing in particular that the energization due to non-resonant interaction with inertial Alfvén waves leads to broadband particle acceleration \cite{damiano_2023}. Resonant wave-particle interaction can occur when the length scale of waves becomes close to the electron inertial scale, allowing an energy transfer between the waves and the particles \cite{saur_2018}, which may happen close to Jupiter, in low-altitude regions.

 Electron and proton broadband spectra have been reported in numerous Juno in situ observations, either in the auroral regions \cite{mauk_2017,allegrini_2020_main_oval} or in the moons' magnetic flux tubes \cite{szalay_2018,szalay_2020a,szalay_2020b,szalay_2020c,clark_2018,clark_2020,rabia_2023}, confirming the prominent role played by Alfvén waves to accelerate particles. Recently, \citeA{sarkango_2024} also reported observations of banded proton and electron distributions within flux tubes connected to the Galilean moons, consistent with a resonant process between particles' bounce motion and standing Alfvén waves. However, \citeA{allegrini_2020b} and \citeA{hue_2022} have shown that electron spectra measured during crossings of two particular flux tubes of Europa and Ganymede, likely connected to their TEB spots, do not always exhibit a broadband distribution but rather a non-monotonic one with a peak in the Differential Number Flux (DNF) spectrum. \citeA{rabia_2023} have demonstrated that, at Europa, a progressive transition occurs as a function of downward distance from the moon, from non-monotonic spectra first observed in the TEB flux tubes to broadband spectra observed farther down the tail. These observations suggest that other physical processes may occur during the electron acceleration in TEB flux tubes leading to modifications of their energy spectra. 

Similarly, observations of UV footprints associated with the moon Enceladus \cite{pryor_2011,pryor_2024} and intense waves activity within its flux tubes \cite{sulaiman_2018} have revealed that similar processes occur during moon-magnetosphere interactions at Saturn. However, the high-latitude properties of the particles accelerated during the Enceladus-magnetosphere interactions remain unknown, preventing any detailed comparisons with observations reported in Jupiter's magnetosphere.

 In this study, we investigate the properties of electrons in Ganymede's flux tubes. We use a cross-correlation between Juno/JADE and Juno/UVS data to identify clear signatures of Ganymede-magnetosphere interactions in the far field regions, i.e., at high latitude close to Jupiter. The instruments, the physical parameters derived from the data, and the methods used in this study are described in Section 2. The statistical properties of electrons in Ganymede's flux tubes are presented in Section 3. Finally, Section 4 and 5 discuss the results in the context of electron acceleration in the Ganymede-magnetosphere interaction, summarize the main conclusions, and offer perspectives for future works.

\section{Methods}

\subsection{Considered data set}

We use in situ measurements made by the JADE-E instrument \cite{mccomas_2017} obtained above Jupiter's polar regions. In these regions, the JADE-E instrument can measure the properties of electrons initially between 50 eV and 100 keV at a 1-second cadence. This energy range was reduced to 32 eV - 32 keV after PeriJove (PJ) 32. Thanks to its two sensors that provide each a 120°-field of view, a large fraction of the pitch angle distribution can be reconstructed, depending on the orientation of the spacecraft with respect to the local magnetic field. Pitch angles are derived using magnetic field directions obtained by the MAG instrument \cite{connerney_2017}. 


The proximity of Ganymede’s footprint with the main auroral oval, and the dynamics observed in the Jovian middle magnetosphere \cite{mauk_2002} can make the identifications of Ganymede’s flux tube crossings in the JADE data sometimes challenging. To overcome this difficulty, we used observations from the Juno/UVS instrument to cross-correlate the in situ and remote observations. UVS is an imaging spectrograph characterizing the jovian auroral emission between 68 and 210 nm \cite{gladstone_2017}, and designed to provide context spectral images to the Juno in situ and remote sensing instruments. During the perijove observation phase, UVS scans the auroral region of Jupiter every 30\,s. By co-adding consecutive images, it is possible to build up complete maps of the auroral regions \cite{bonfond_gladstone_2017, greathouse_2021}. Because of the short lifetime of Jupiter's auroral structures, comparisons between the Juno in situ measurements and UVS imagery must be done with data recorded simultaneously. 

IR images and spectra of the Ganymede footprint, taken from Juno, exist as well \cite{mura_2017,mura_2018}. However, the typical H$_{\mathrm{3}}^{+}$ decay time, which is responsible for IR emission, is variable and can be between tens to a few hundred seconds. This causes the  morphology of UV and IR footprints to be different \cite{gerard_2018} and also makes instantaneous comparison of remote and in situ data impossible. In addition, the pointing flexibility offered by UVS scan mirror allows targeting specific auroral features, making the UVS satellite auroral footprint dataset extensive \cite<e.g.,>{hue_2019,hue_2023}. For this reason, we focused our study on the UV images of the Ganymede footprint. 

We used here UV-false color images derived by spectrally binning the UVS data into three `color' channels. The red, green and blue channels account for the photons recorded by UVS in the 145-165\,nm, 119-145\,nm, and 70-119\,nm ranges, respectively. The intensity of the red channel is a proxy for the electron characteristic energy, in a similar fashion as the traditional color ratio \cite{Yung1982}. That quantity uses the fact that the absorption cross-section of methane drops off sharply at wavelengths longer than 145\,nm. The absorption of $<$ 145\,nm UV auroral emissions from stratospheric methane increases with the depth of precipitating electrons, and therefore with the electron characteristic energy. Comparing the false color of the auroral features observed at the Juno footprint provides an additional way to correlated in situ with remote sensing measurements \cite{szalay_2020a}, instead of only looking at the Juno calculated M-shell from the magnetic mapping

The satellite footprints result from the interaction between the co-rotating plasma and the moons. These spots therefore do not co-rotate with the planet and can be distinguished from other auroral emissions as their rotation rate follows the orbital motion of the moons. UVS spectral images were produced by co-adding a number of spins around the Ganymede flux tube crossing times \cite<see, \textit{e.g.}, the method used in>{hue_2022}. The UVS time window was optimized per case to produce an image co-added over a large-enough time window providing a complete picture of the auroral emissions near the footprints, but short enough to limit the smearing of the footprints.

\subsection{Identification of the crossings of the Ganymede flux tubes}

In order to identify potential crossings of Ganymede's flux tubes by Juno in the JADE data obtained between PJ1 and PJ50, we apply the step-by-step method described hereafter. 

(i) We first rely on a magnetic field lines tracing method to estimate when Juno magnetic footprint crossed the Ganymede footpath, i.e., the location on the Jupiter poles of the magnetic field lines connected to Ganymede's orbital location. For this purpose, we used the JRM33 magnetic field model \cite{connerney_2022} coupled to the CON2020 current sheet model \cite{connerney_2020} since this combination gives the most accurate description of the magnetic field at the orbital locations of Io, Europa, and Ganymede \cite{rabia_2024}. The magnetic field lines are iteratively traced with a constant step size of 1/350 $\rj$ (1 $\rj$ = 71,492 km), i.e. $\sim$ 200 km, toward the auroral regions. We projected the position of Juno onto Jupiter's atmosphere at a 900 km-altitude, which corresponds to the mean altitude of the moon-induced UV aurora \cite{bonfond_2009}, and estimated the times when it crossed the predicted footpath of Ganymede. Altitudes are calculated above the 1-bar level. The footpaths of the moons considered, also derived using JRM33+CON2020 model, are available in \citeA{connerney_2022}. A catalog of $\sim$ 150 potential crossings is built, covering the Juno orbits between PJ1 and PJ50, corresponding to $\sim$ 3 crossings per orbit. 

(ii) We restrict the catalog to the cases where Juno's altitude is below 1.5 $\rj$ to ensure that the loss cone (LC) is sufficiently large to be resolved by the JADE-E instrument. Below this altitude, the LC is larger than 15$^{\circ}$. This criterion also ensures that the Juno altitude is small enough to measure accelerated electrons, i.e., that Juno's is below or within the acceleration region. This criterion excludes 40 cases of the catalog, 110 potential crossings of Ganymede's flux tubes remain.  

(iii) We limit our catalog to the cases where Juno crossed the Ganymede footpath close to the Ganymede magnetic footprint. For this purpose, we add a criterion on the longitude separation between the magnetic footprint of Juno and the location of Ganymede MAW spot, i.e., -20° $< \dalfven <$ 60° (see Section 2.3 for the definition of the $\dalfven$ metric), to ensure that electrons fluxes resulting from the Ganymede-magnetosphere interactions are sufficiently strong to be measured in situ above the background magnetospheric particle flux. Indeed, \citeA{bonfond_2017} estimated that Ganymede's UV tail has a length of $\sim$ 24°, which provides a typical spatial scale over which the electron fluxes decreases. 23 potential crossings remains after this stage. 

(iv) We analyze the UVS data for the 23 potential crossings identified. We discard the cases in which the Ganymede UV footprint and auroral tail are not clearly visible on the Jupiter pole and/or not distinguishable from nearby auroral emissions unrelated to satellite auroral emissions. We then analyze the time series of JADE-E measurements and correlate the in situ electron observations with their auroral counterparts observed by UVS along the footpath of Juno, as presented in Figure 1 and Figure 2. By correlating the sequence of events observed in both instruments (see Section 3.1), we obtain 11 cases where we observe electrons accelerated by the Ganymede-magnetosphere interactions. A similar procedure was also used by \citeA{allegrini_2020b} and \citeA{szalay_2020a} to identify a Europa footprint crossing during PJ12N and a Ganymede MAW flux tube crossing during PJ20N, respectively. We emphasize that the 11 events reported in this study are the only ones to exhibit a clear acceleration signature in the JADE-E data.

\subsection{Electron properties characterization}

We characterize the electron properties by deriving the precipitating energy flux ($EF$) and the precipitating electron characteristic energy ($\ecar$) using the formulas presented in \citeA{rabia_2023}. All the quantities presented hereafter are made using electron measurements within the loss cone, whose size is estimated based on the formula of \citeA{mauk_2017}. 

In Section 3, the 11 crossings of the Ganymede's flux tubes identified are organized according to the $\dalfven$ separation, the angle along each moons' orbit between the moon and an Alfvén wave trajectory connected to the tail aurora \cite{szalay_2020b}, and where $\dalfven$ = 0° theoretically corresponds to the MAW spot. This metric takes into account the propagation of Alfvén waves from the close environment of the moons up to Jupiter's ionosphere and is based on an empirical model derived from UVS observations of the Io, Europa, and Ganymede MAW spots \cite{hue_2023}. By providing an estimate of the downtail separation from the MAW spot, previous studies at Io and Europa have shown that the $\dalfven$ metric is the most suitable parameter for organizing electron fluxes as it allows to minimize the scatter of the $EF$ downtail of the MAW \cite{szalay_2020b, rabia_2023}.

This $\dalfven$ separation allows to divide the observations into three regions depending on the $\dalfven$ separation, i.e., TEB for $\dalfven$ $<$ -3°, MAW for -3° $<$  $\dalfven$ $<$  3°, and auroral tail for $\dalfven$ $>$  3°. Indeed, since the $\dalfven$ metric quantifies the angular separation from the MAW spot, a negative $\dalfven$ value likely refers to the crossing of a structure upstream of the MAW, most likely the TEB. However, as this metric can be greatly affected by the time variability of the magnetospheric conditions \cite<see>{hue_2023}, e.g., plasma density, magnetic field strength, magnetodisc scale height, we consider that the crossing occurring at $\dalfven$ = $\pm$ 3° can still be related to the MAW. For larger downtail separations, the flux tubes crossed are connected to Ganymede's auroral tail. 

\section{Results}
\subsection{Ganymede-magnetosphere interaction signatures in JADE and UVS data}

\begin{figure}[h!]
    \hspace{-2cm}
    \includegraphics[width=1.3\textwidth]{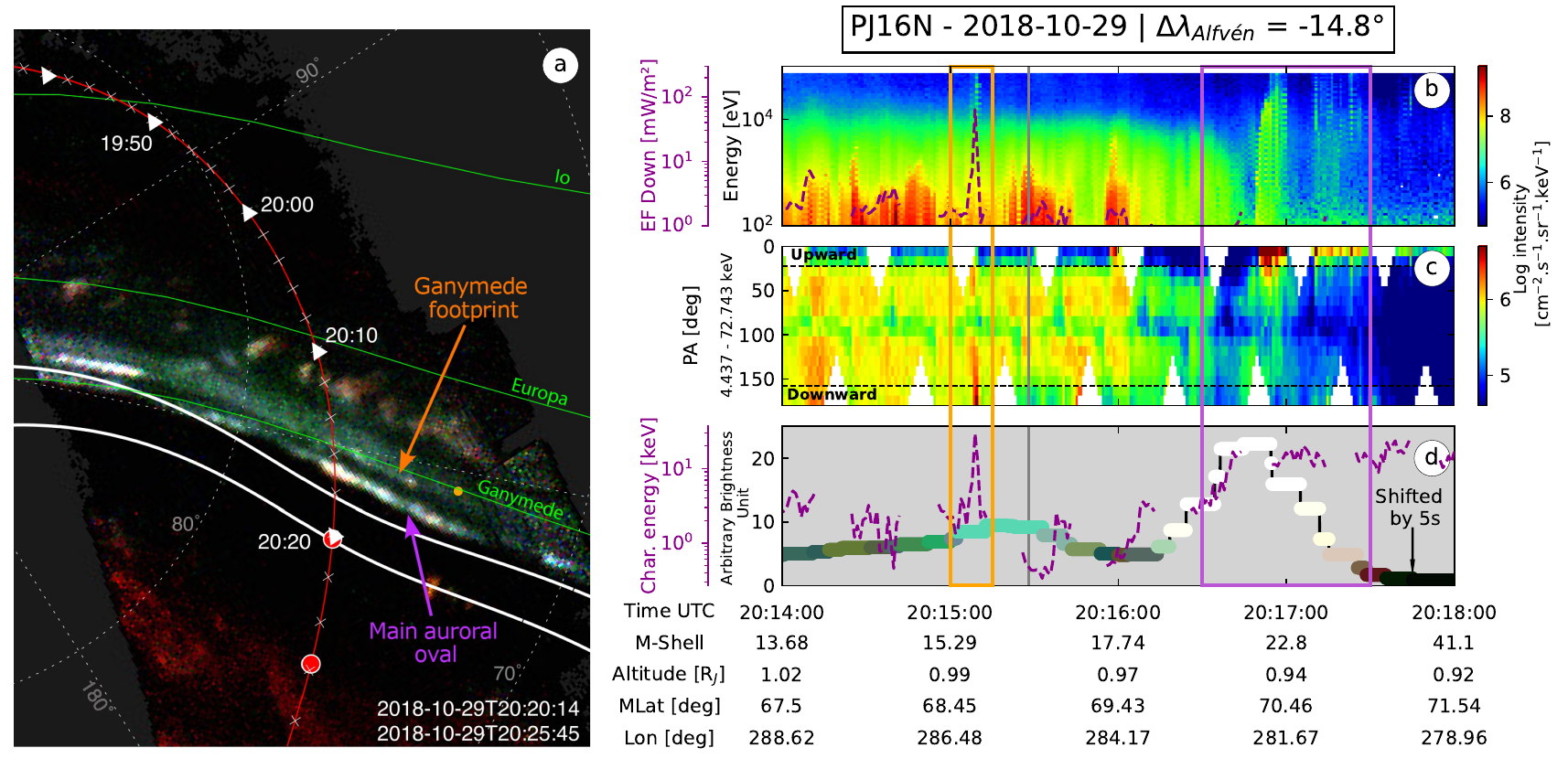}
    \caption{JADE in situ and UVS remote observations made during PJ16N. (a) UV false color image of the auroral emissions observed close to Ganymede’s footprint. The Juno footpath and the instantaneous footprints of the Galilean moons derived using the JRM33+CON2020 models are shown as red lines and orange dots, respectively. The boundary of the time window over which the UVS data has been co-added is shown at the bottom right corner, and is reported on the Juno footpath as two red dots. The white triangles and crosses along the Juno footpath highlight Juno's instantaneous footprints with a 10 and 2 minutes time step, respectively. (b) Electron differential number flux measured by JADE-E. The $EF$ is superimposed. (c) Pitch angle distribution. The size of the loss cones at the time of the flux tube crossing is indicated by the dashed line. (d) Arbitrary brightness related to the number of UVS total countrate, UVS false color along the footpath of Juno and electron characteristic energy. The information related to the position of Juno are indicated below the panel (d). The times for the flux tubes crossings estimated from JRM33+CON2020 are represented by vertical gray lines in panels (b, c, d). Events associated with a Ganymede's flux tube crossing are highlighted by an orange rectangle while the crossing of the main oval is displayed, in panels (b, c, d), as a purple rectangle.}
    \label{pj16}
\end{figure}

\begin{figure}[h!]
    \hspace{-2cm}
    \includegraphics[width=1.3\linewidth]{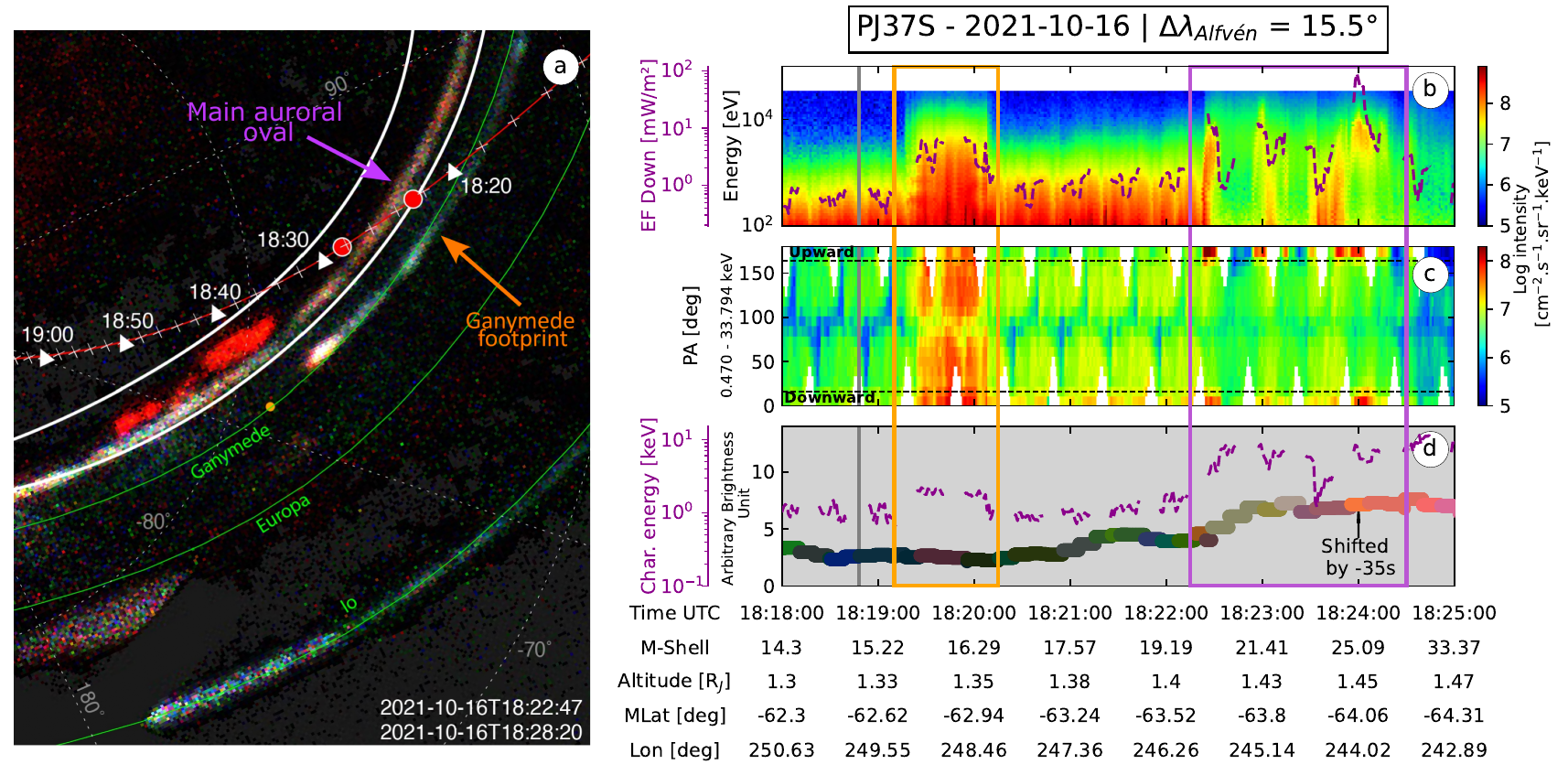}
    \caption{Same as Figure 1, for PJ37S}
    \label{pj37}
\end{figure}

Figure \ref{pj16} and Figure \ref{pj37} show the JADE and UVS data obtained during PJ16N and PJ37S, respectively. During these flux tube crossings, the $\dalfven$ separations are -14.8° and 15.5°, suggesting a TEB and a tail crossing, respectively. The observations made during the other events studied, not shown in Figure \ref{pj16} and Figure \ref{pj37}, are displayed in the Supplementary Information file. Table \ref{table1} also summarizes the times, positions, and electron properties of all 11 events.

Figures \ref{pj16}a and \ref{pj37}a show the auroral structures observed by UVS during PJ16N and PJ37S, as well as Juno's magnetic footpath projected onto Jupiter's atmosphere, shown as red lines and dots. During the time ranges represented, Juno crossed magnetic field lines with increasing M-Shell values. As a result, the spacecraft first crossed the footpath of Io (PJ16N $\sim$ 19:46, not shown for PJ37S), followed by the one of Europa (PJ16N $\sim$ 20:09, not shown for PJ37S), and the one of Ganymede (PJ16N $\sim$ 20:15, PJ37S $\sim$ 18:19). Finally, Juno crossed the main oval emission region few minutes after it crossed the footpath of Ganymede. The measurements represented in Figure \ref{pj16} (b, c, d) and Figure \ref{pj37} (b, c, d) are obtained close to the estimated time of the crossings of Ganymede's footpath. The crossings of the main oval are identified as large increases of the $EF$, $\ecar$, and auroral brightness \cite{allegrini_2020_main_oval,mauk_2017,clark_2018}. Few minutes before the crossings of the main oval, a sudden increase by a factor of $\sim$ 10 in the electron differential number flux is observed, at a time close to the one estimated by JRM33+CON2020 for each crossings, that we identify as a signature of the Ganymede-magnetosphere interaction. This conclusion is reinforced by the M-Shell values crossed by Juno at the time of the observations, i.e., 15.5 and 15.9 for PJ16N and PJ37S, respectively. These values fall within the M-Shell range covered by Ganymede, from 15 to 17, because of the tilt between the magnetic equator and the orbital plane of the satellites. We note that for the 11 events reported, the median time deviation between the JRM33+CON2020 estimates of the flux tubes crossings and the actual start of the beginning of the crossings as observed in the JADE data is 13 seconds. This value, due to magnetic field model and mapping uncertainties, is close to the 14 seconds median deviation derived by \citeA{rabia_2024} who compared the JRM33+CON2020 estimates with crossings of Io, Europa, and Ganymede flux tubes with in situ observations reported in several publications. In this study, we add 9 potential observations of Ganymede's flux tubes crossings not considered in \citeA{rabia_2024}, reinforcing the conclusion that JRM33+CON2020 is accurate in predicting the crossings of moons' magnetic flux tubes.

\begin{figure}[h!]
    \hspace{-2cm}
    \includegraphics[width=1.3\linewidth]{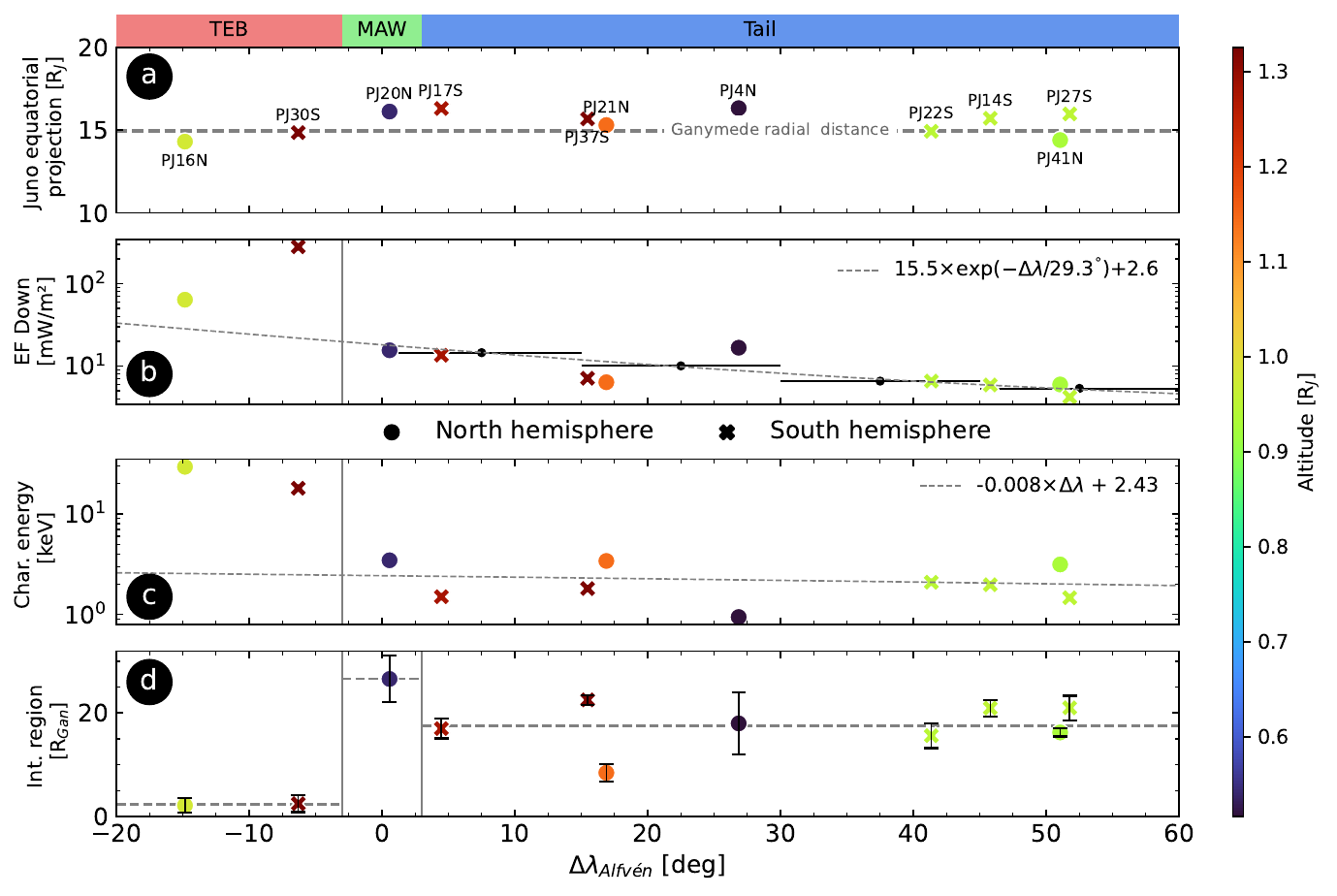}
    \caption{(a) Projection of Juno position onto Ganymede’s orbital plane at the times of flux tubes crossings. Ganymede orbital distance is indicated with a gray dotted line. (b) Precipitating $EF$. The exponential decay law that best fits the dataset is shown. Horizontal black lines represent the width of the bins used to bin the energy flux values. (c) Characteristic electron energy. (d) Size of the interaction region projected onto Ganymede orbital plane and estimated from the duration of the acceleration events. The color code used refers to the altitude of the crossing while the markers indicate the hemisphere in which the flux tube crossing was observed. At the top of the figure, we indicate the auroral structure created by the electrons within the crossed flux tubes, estimated with the $\dalfven$ separation.}
    \label{stastique}
\end{figure}

The signatures of the Ganymede-magnetosphere interactions appear as sudden increases of the electron differential flux, with a typical duration of few tens of seconds (Figures \ref{pj16}b and \ref{pj37}b). This duration strongly depends on the configuration of the flux tube crossing. When simultaneous coverage of the upward and downward LC is available, the observed electron beams are always bidirectional in the pitch angle domain (Figures \ref{pj16}c and \ref{pj37}c), which is consistent with the previous observations of the Io- and Europa-magnetosphere interactions \cite{szalay_2018,szalay_2020a,szalay_2020b,allegrini_2020b, rabia_2023, hue_2022}. This feature, predicted by theoretical \cite{hess_2010} and modeling \cite{damiano_2019} works, suggests an acceleration by inertial Alfvén waves. The observation of both downward and upward filled loss cones also suggests that Juno lies within the acceleration region. Indeed, only the downward LC is expected to be filled if Juno is below the acceleration region, and that only the upward one if Juno is above this region. The altitude range sampled during the observations reported, i.e., 0.5 - 1.3 $\rj$ (Figure \ref{stastique}, Table \ref{table1}) could therefore be an indication of the altitude of the acceleration region. 
 
During PJ16N, identified as a TEB crossing, the electron characteristic energy $\ecar$ raises up to $\sim$ 30 keV while during PJ37S, identified as a tail crossing, this increase is smaller, around 2 keV. These values are representative of the other events studied, with a significantly higher $\ecar$ in the TEB compared to the one in the MAW and tail (Table \ref{table1}, Figures \ref{stastique}c).  

Figures \ref{pj16}d and \ref{pj37}d show the brightness and the false color UV emissions observed along the Juno footpath during PJ16N and PJ37S, respectively. For both events, we note a small time offset between the JADE and UVS measurements. Indeed, taking the PJ37S observations as an example, the crossing of the main oval is observed in the JADE-E measurements at around 18:22:30 while it appears in the UVS data as a brighter emission along Juno's footpath at around 18:23:30. Similar time offsets were also observed by \citeA{allegrini_2020_main_oval} and \citeA{al_saati_2022} who studied the crossings of the main auroral oval during the first 9 and 30 perijoves, respectively. This may result from (i) uncertainties on the projections of the Juno magnetic footprint or the auroral emissions onto Jupiter's poles, independent from the magnetic mapping uncertainties (ii) temporal differences due to the fact that UVS observations are not truly simultaneous with JADE and instead results from integrating over a certain temporal range as close as possible from the JADE measurement window, and/or (iii) projection effects resulting from UVS observing extended auroral structures smeared over several co-added Juno spins with different vantage points. We quantify the JADE-UVS temporal offset by comparing their observations of the main oval crossings for all the events studied. We note that this quantification is only an estimate of the JADE-UVS temporal offset since it remains challenging to identify precisely the start of the main auroral oval crossing in both dataset \cite{allegrini_2020_main_oval}. The values derived are provided in Table 1. They correspond to the times we applied to the UVS data to match the JADE-E in situ measurements. 


\begin{table}[h!]
\caption{Times and positions for which we identified Ganymede-magnetosphere interaction signatures in the JADE-E and UVS data. Electrons properties derived, displayed in Figures \ref{pj16}, \ref{pj37} and \ref{stastique} are reported.}
\label{table1}
\resizebox{1.1\textwidth}{!}{\hspace{-1.75cm}
\begin{tabular}{ccccccccccccc}
PJ   \& Hem & 
Date & 
T$_{\mathrm{start}}$ &
T$_{\mathrm{stop}}$ & 
T$_{\mathrm{predicted}}$ & 
\begin{tabular}[c]{@{}c@{}} $\dalfven$\  \\  {[}°{]}\end{tabular} & 
\begin{tabular}[c]{@{}c@{}}Juno alt \\ {[}$\rj${]}\end{tabular} & 
\begin{tabular}[c]{@{}c@{}}Gan. $\theta_{\mathrm{mag}}$ \\ {[}°{]}\end{tabular} & 
\begin{tabular}[c]{@{}c@{}} $EF$ \\ {[}mW/m²{]}\end{tabular} & 
\begin{tabular}[c]{@{}c@{}} $\ecar$\\  {[}keV{]}\end{tabular} & 
\begin{tabular}[c]{@{}c@{}}Interaction region \\ size {[}$\rgan${]}\end{tabular}  & 
\begin{tabular}[c]{@{}c@{}}UVS integration \\ time   {[}s{]}\end{tabular} & 
\begin{tabular}[c]{@{}c@{}}UVS-JADE\\  shift {[}s{]}\end{tabular} \\ \cline{1-13}
16N & 2018-10-29 & 20:15:06 & 20:15:09 & 20:15:28 & -14.83 & 0.980 & -8.86 & 63.88  & 29.46  & 2.112  & 331 & 5   \\
30S & 2020-11-08 & 02:55:00 & 02:55:03 & 02:54:58 & -6.30  & 1.325 & 1.82  & 280.0  & 18.034 & 2.435  & 572 & 10   \\
20N & 2019-05-29 & 07:37:19 & 07:37:31 & 07:37:36 & 0.57   & 0.540 & -8.35 & 15.524 & 3.459  & 26.598 & 457 & -3  \\
17S & 2018-12-21 & 18:03:40 & 18:03:58 & 18:03:06 & 4.47   & 1.279 & 6.94  & 13.539 & 1.5    & 16.999 & 1548 & 30  \\
37S & 2021-10-16 & 18:19:18 & 18:20:07 & 18:18:48 & 15.48  & 1.322 & 8.06  & 7.0869 & 1.811  & 22.543 & 333 & -35 \\
21N & 2019-07-21 & 03:05:24 & 03:05:34 & 03:05:14 & 16.89  & 1.144 & -1.38 & 6.349  & 3.407  & 8.434  & 452 & 20  \\
4N  & 2017-02-02 & 12:26:23 & 12:26:29 & 12:26:36 & 26.85  & 0.516 & -5.99 & 16.684 & 0.942  & 17.997 & 638 & 5   \\
22S & 2019-09-12 & 04:30:18 & 04:30:31 & 04:30:22 & 41.35  & 0.949 & -7.57 & 6.553  & 2.091  & 15.611 & 632 & 35  \\
14S & 2018-07-16 & 06:07:30 & 06:07:57 & 06:07:30 & 45.79  & 0.954 & 10.11 & 5.88   & 1.978  & 20.944 & 992 & 5   \\
41N & 2022-04-09 & 14:59:18 & 14:59:59 & 15:00:22 & 51.05  & 0.930 & -2.85 & 5.967  & 3.143  & 16.243 & 1160 & -10 \\
27S & 2020-06-02 & 11:10:13 & 11:10:31 & 11:10:00 & 51.77  & 0.950 & -0.45 & 4.19   & 1.466  & 21.014 & 1053 & -50
\end{tabular}%
}
\end{table}

\subsection{Statistical study}

In order to confirm that the observations are related to a Ganymede's flux tube crossing by Juno, we project the spacecraft position at the time of the crossing onto the moon's orbital plane. The results, presented in Figure 3a, show that all the signatures observed in the JADE-E data map between R = 14.3 $\rj$ and R = 16.3 $\rj$, which is close to Ganymede's orbital distance, i.e., R = 14.95 $\pm$ 0.02 $\rj$. Deviations from Ganymede's semi-major axis may result from (i) magnetic field mapping uncertainties or (ii) a non-stricly azimuthal translation of the disturbances in the wake of Ganymede, which modified the radial distance location of the electron beams source.

Figure 3b shows the maximum precipitating $EF$, i.e., the $EF$ within the downward LC, measured during each of the flux tubes crossings identified as a function of the $\dalfven$ separation. We also bin the $EF$ values in 15° intervals and perform an exponential fit, excluding the $EF$ measured in the TEB. We find that the $EF$ decreases with downtail distance. In Figure 3b, we show the decay profile that minimizes the root mean square error (RMSE = 0.235). We find that the e-folding value of $\sim$ 29°, which quantifies the $EF$ decrease with the downtail distance, better matches our limited data set. This is consistent with the e-folding value of 24° derived by \citeA{bonfond_2017} that fit the decrease of the Ganymede's auroral tail UV brightness using HST observations. The fit presented in Figure 3c, which also minimize the RMSE (RMSE = 0.852), shows that the correlation between $\ecar$ and $\dalfven$ is very weak, meaning that $\ecar$ is nearly constant in the MAW and auroral tail flux tube.  


In Figure 3d, we estimate the size of the interaction region in the equatorial plane in these three regions. To do so, we project Juno's position at the start and end of the flux tubes crossings onto Ganymede's orbital plane. Error bars are derived assuming a $\pm$ 1 second uncertainty on the start and end of the flux tubes crossing as observed in JADE-E data. We observe a large discrepancy between the size of the interaction region derived in the TEB, MAW, and auroral tail. Using the two crossings of the TEB, we derive a mean size for the interaction region of 2.3 $\pm$ 1.5 $\rgan$, where 1 $\rgan$ = 2,634 km. The size obtained using with the MAW crossing, i.e., $\sim$ 26.5 $\pm$ 4.4 $\rgan$, is the largest. Finally, in the tail, the mean size derived by averaging 8 observations at various $\dalfven$ separations is $\sim$ 17.5 $\pm$ 2.2 $\rgan$. With our limited data set, we do not find any correlation between the interaction size and $\dalfven$ separation as reported for Io \cite{szalay_2020b} but rather a constant value as reported for Europa \cite{rabia_2023}.


\subsection{Energy distribution}

 We have shown that Juno sampled electrons within Ganymede's TEB, MAW, and auroral tail. Here, we investigate their properties in these different regions, in particular their energy spectra. Figure 4 shows the electron energy distribution, i.e., the electron DNF within the downward LC as a function of energy, for PJ16N, PJ20N, and PJ37S which are identified as crossings of the TEB, MAW and auroral tail, respectively. The electron spectra for the other events studied, not presented here, are provided in Figure S6.  

\begin{figure}[h!]
    \hspace{-3cm}
    \includegraphics[width=1.5\linewidth]{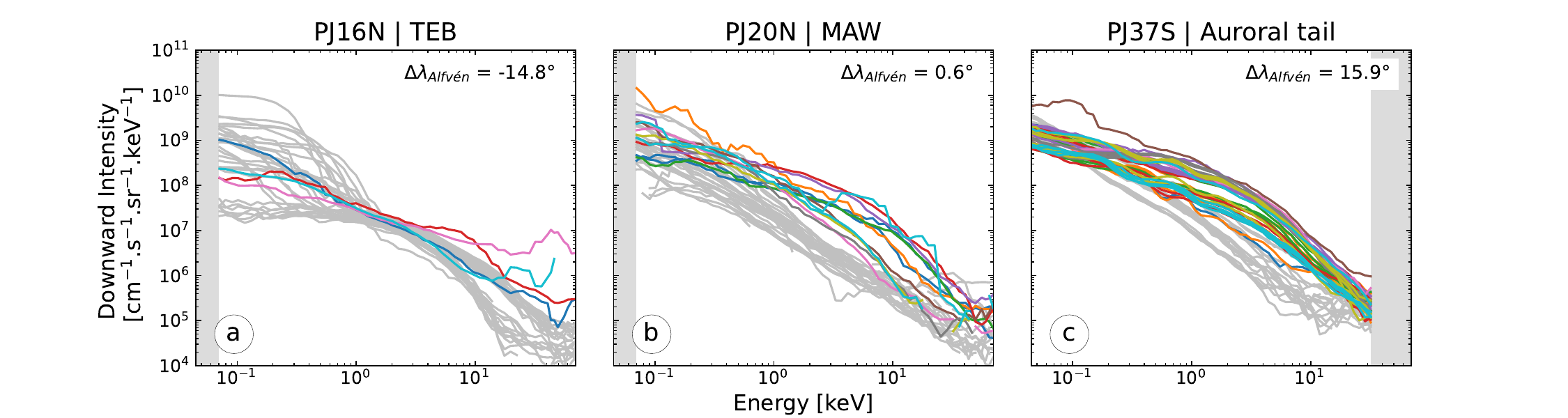}
    \caption{Electron energy distribution within the downward LC observed in flux tubes connected to the TEB spot (a), MAW spot (b) and auroral tail (c). Colored curves show the distribution every second measured within the flux tubes. Gray curves represent the distribution measured in 15-second intervals before and after the flux tube crossing. Gray areas indicate the energy range not covered by the JADE-E instrument as a result of a change of in instrument operation after PJ32.}
    \label{energy_distribution}
\end{figure}

It appears that in the TEB (Figure 4a), the electrons exhibit non-monotonic spectra, with a clear increase of the DNF at a specific energy which correspond roughly to $\ecar$. At others energies, the DNF is similar to the magnetospheric background flux. The other Ganymede's TEB crossing observation, reported by \citeA{hue_2022} presented a similar behavior, with a DNF increase for electrons in the 2-35\,keV range. We note that the electrons in Europa's TEB identified by \citeA{rabia_2023} also exhibit a non-monotonic spectra, albeit with a lower $\ecar$.  

In the MAW and tail, the electron energy distributions take the form of broadband spectra (Figure 4b and 4c). The electron DNF increase is observed over a wide energy range, i.e., for E$>$ 1 keV in the MAW (Figure 4b) and E $>$ 0.1 keV in the auroral tail (Figure 4c). This means that electrons are accelerated at all the corresponding energies. This type of distribution has been systematically observed in flux tubes connected to Io, Europa, and Ganymede MAW and auroral tails \cite{szalay_2020a,szalay_2020b,rabia_2023}.

\section{Discussion}

Using a cross-correlation between magnetic field line mapping, JADE, and UVS data, we identified 11 events during which Juno crossed the magnetic shell of Ganymede. For these events, Juno observed bidirectional electrons beams (Figures 1c and 2c), with a characteristic energy between 1 keV to 30 keV (Figure 3c). These measurements suggest that Juno transited within the acceleration region, at an altitude between 0.5 and 1.3 $\rj$. 


We organized the observations as a function of the $\dalfven$ separation, a metric that takes into account the propagation of Alfvén waves from moons' orbits to Jupiter's atmosphere based on HST or Juno-UVS observations \cite{hue_2023}, as previously done for Io and Europa studies \cite{szalay_2020b,rabia_2023}. We show that for Ganymede the $\dalfven$ metric is also the most appropriate one to organize the EF as a function of downtail distance (Figure 3b). This does not rule out that a dependence with Ganymede's magnetic and centrifugal latitudes could exist too since it is expected that Ganymede encounters a denser plasma flow when the moon lies within the magnetodisc. Such a dependency may however be difficult to identify since the time variability of the magnetospheric conditions also plays a role in the strength of the moon-magnetosphere interactions. In contrast, there is no clear correlation between $\ecar$ and the $\dalfven$ separation, $\ecar$ being almost constant in the MAW and auroral tail, with a mean value of 2.2 keV.  

Following the observation of a correlation between the precipitating $EF$ and the $\dalfven$ separation, we constrain an e-folding value of 29° for the in situ measurements (Figure 3b), consistent with 24° derived from fitting the brightness decay of Ganymede's auroral tail using HST observations \cite{bonfond_2017}. We note that previous statistical studies on Io and Europa derived $EF$ e-folding values of 21° and 7°, respectively \cite{szalay_2020b,rabia_2023}. It is surprising that the e-folding value of electron $EF$ at Ganymede is slightly higher than the one of Io as the extent of Io's auroral tail extent is larger than that of Ganymede. A greater number of Ganymede flux tube crossings would help to better constrain the $EF$ e-folding. 

We found a clear distinction between the properties of the electrons creating the TEB spot and those measured in the MAW, and auroral tail flux tubes. Indeed, it appears that electrons measured within the flux tubes connected to Ganymede's TEB spot ($\dalfven$ $<$ -3°) exhibit a significantly higher downward $EF$ and $\ecar$ (Figure 3b, 3c). In Europa's TEB flux tubes, higher $EF$ and $\ecar$ were also reported \cite{rabia_2023}. At Io, an observation of a large $EF$ of $\sim$ 600 mW/m² has been reported by \citeA{szalay_2020b} but the positive $\dalfven$ separation during this event was not consistent with a TEB crossing which was therefore attributed to a MAW crossing. 

We estimated the size of the interaction region in Ganymede's orbital plane by mapping magnetically the position of Juno during the crossings. We found a clear distinction between the estimates based on TEB crossings and those derived from crossings of flux tubes connected to the auroral tail. By expressing the size estimated using TEB crossings in kilometers and comparing to the one reported at Europa \cite{allegrini_2020b,rabia_2023}, we found a size of $\sim$ 6000 km for both moons. These similar observations, made on two moons of very different sizes, may suggest that the size of the interaction region derived using TEB crossings are independent of the size of the obstacle, i.e., the moons, but are rather an indication of the width of the flux tubes where the electron are accelerated in the upward direction. However, the low number of TEB crossing observations does not allow us to drawn firm conclusion about the origin of this smaller interaction size derived. The size derived using the MAW crossing during PJ20N, i.e., 26.5 $\pm$ 4.4 $\rgan$, is slightly higher than the estimate of 8-20 $\rgan$ estimated from the size of the MAW spot observed by HST \cite{grodent_2009}. In the tail, we estimated the size of the interaction region to be 17.5 $\pm$ 2.2 $\rgan$, which is $\sim$3-4 times larger than the estimates at Io and Europa (5.6 $\rio$ and 4.2 $\reur$). We note that estimates of the interaction region size larger than the size of the moons have been also reported at Io and Europa \cite{szalay_2020b,rabia_2023}. However, this must be compared with the size of the obstacle offered to the magnetospheric plasma flow by these three objects. Considering that Io interacts with the extent of its ionosphere \cite<$\sim$ 2.8 $\rio$, >{saur_2021_overview}, Europa only with its small exosphere ($\sim$ 2 $\reur$) and Ganymede with its magnetosphere \cite<$\sim$ 6 $\rgan$, >{jia_2009,jia_2021}, a correlation between the size of the obstacle $R_{\mathrm{obstacle}}$ and the size of the interaction region in moons' wake derived with flux tubes crossings $R_{\mathrm{wake}}$ can be established. Based on the values reported at Io and Europa by \citeA{szalay_2020b} and \citeA{rabia_2023}, we find that ($R_{\mathrm{wake}}$/$R_{\mathrm{obstacle}}$) = 2 at Io, ($R_{\mathrm{wake}}$/$R_{\mathrm{obstacle}}$) = 2.1 at Europa and ($R_{\mathrm{wake}}$/$R_{\mathrm{obstacle}}$) = 2.9 at Ganymede. These observations, which do not reflect the results of MHD simulations of the local environment of the moons, require investigation by future studies considering the propagation of the disturbances in far-field regions.

These two different regimes are also found when looking at the electron energy distributions. In the MAW and auroral tail, the electron distributions are always observed to be broadband. These broadband distributions could be explained by acceleration processes involving inertial Alfvén waves \cite{damiano_2019}. Evidence for Alfvénic acceleration in the flux tubes of Ganymede has also been raised by \citeA{Louis_2020,louis_2023} and \citeA{mauduit_2023} who use a diagnostic of the Ganymede-induced radio emission. Since the TEB is formed by an anti-planetward acceleration from the MAW \cite{bonfond_2008}, we expect to observe the same distribution for electrons creating the MAW and TEB spots. However, we have shown that in the TEB, electrons exhibit a non-monotonic distribution with a DNF increase at a specific energy close to $\ecar$ (Figure 4a). We suggest that only a part of an initial broadband energy distribution resulting from an anti-planetward acceleration reaches the opposite hemisphere. During the propagation from the acceleration region toward the opposite hemisphere, most of the energy distribution might be scattered and/or undergoes additional wave-particle interactions occurring during the crossing of the magnetodisc, preventing them to propagate all the way down to the auroral regions. The energy distribution measured in the TEB flux tube, which only contains the particles that propagate all the way, is therefore different from the initial broadband distribution. 

Since we only reported in the present paper observations of electrons within the downward loss cone, additional studies are required to confirm that the electron energy distribution in the upward loss cone is also broadband and compatible with an acceleration process involving Alfvén waves. Moreover, measurements during a crossing of Io's TEB, still unreported, could provide new insights into the origin of the TEB in moon-magnetosphere interactions.

\section{Conclusions}

We characterized the properties of electrons accelerated by the Ganymede-magnetosphere interactions using a joint analysis of Juno in situ plasma and remote UV auroral measurements. The main outcomes of our statistical analysis are as follows:  

\begin{enumerate}
    \item The combination of JADE-E measurements and UVS observations with field line tracing enables us to identify 11 events when Juno can diagnose in situ the properties of electrons accelerated by the Ganymede – Jovian magnetosphere interaction.
    \item We find a clear dependency of $EF$ with the $\dalfven$ separation. The e-folding value from our in situ observations, i.e., 29°, is consistent with the value of 24°, previously derived with HST observations. 
    \item The properties of electrons measured in flux tubes connected to the TEB spot, i.e., $EF$, $\ecar$ and energy distribution, are significantly different than those creating the MAW spot and the auroral tail. Further theoretical and modeling studies on acceleration processes are required to explain these differences.  
    \item The size of the interaction region in Ganymede's orbital plane is estimated to be 17.5 $\pm$ 2.2 $\rgan$, based on the measurements made within the flux tubes connected to the auroral tail.
    \item The observation of bidirectional electron beams suggests that Juno was located within the acceleration region during these measurements. Thus, we estimate that this one is located between an altitude of 0.5 $\rj$ to 1.3 $\rj$ at least.   
\end{enumerate}

\section*{Open Research}

JADE \cite{juno_jade_data} and UVS data \cite{juno_uvs_data} used in this study are publicly available on the Planetary Data System, nodes Planetary Plasma Interactions and Planetary Atmospheres, respectively. Part of the analysis have been done using the CLWEB software (\textcolor{blue}{\url{http://clweb.irap.omp.eu}}) developed by E. Penou at IRAP and the AMDA software (\textcolor{blue}{https://amda.cdpp.eu}) provided by CDPP (\textcolor{blue}{http://www.cdpp.eu/}).  


\acknowledgments

French co-authors acknowledge the support of CNES to the Juno mission.  This study has been partially supported through the grant EUR TESS N°ANR-18-EURE-0018 in the framework of the Programme des Investissements d'Avenir. The research performed by N.A. and Q.N. holds as part of the project FACOM (ANR-22-CE49-0005-01 ACT) and has benefited from a funding provided by l’Agence Nationale de la Recherche (ANR) under the Generic Call for Proposals 2022. The work at SwRI was funded by the NASA New Frontiers Program for Juno through contract NNM06AA75C. V.H. acknowledges support from the French government under the France 2030 investment plan, as part of the Initiative d’Excellence d’Aix-Marseille Université – A*MIDEX AMX-22-CPJ-04.


%
%



\bibliography{agusample.bib}

%
%
%
%
%

\end{document}